\documentclass{amsart}
\usepackage{amsmath}
\usepackage{amsfonts}

\theoremstyle{plain}
\newtheorem{theorem}{Theorem}[section]

\theoremstyle{remark}
\newtheorem{remark}[theorem]{Remark}
\numberwithin{equation}{section}

\input{tcilatex}

\begin{document}

\begin{center}
THE\ VACUUM\ STRUCTURE, SPECIAL\ RELATIVITY AND QUANTUM MECHANICS REVISITED:
\ A FIELD THEORY NO-GEOMETRY APPROACH WITHIN THE LAGRANGIAN AND HAMILTONIAN
FORMALISMS. Part 2

\bigskip \bigskip

\bigskip \vspace{1.5cm} Nikolai N. Bogolubov (Jr.)\footnote{%
nikolai\_bogolubov@hotmail.com}\\[0pt]
\bigskip \textit{The V.A. Steklov Mathematical Institute of RAN, Moscow,
Russian Federation \\[0pt]
and \\[0pt]
The Abdus Salam International Centre for Theoretical Physics, Trieste, Italy,%
}\\[1em]
\smallskip Anatoliy K. Prykarpatsky\footnote{%
pryk.anat@ua.fm, prykanat@cybergal.com}\\[0pt]

\bigskip

\textit{The AGH University of Science and Technology, Krak\'{o}w 30-059,
Poland,\\[0pt]
and\\[0pt]
The Ivan Franko State Pedagogical University, Drohobych, Lviv region,
Ukraine \\[0pt]
}

\bigskip
\end{center}

\centerline{\bf Abstract} The work is devoted to studying the vacuum
structure, special relativity, electrodynamics of interacting charged point
particles and quantum mechanics, and is a continuation of \cite{BPT,BPT1}.
Based on the vacuum field theory no-geometry approach, the Lagrangian and
Hamiltonian reformulation of some alternative classical electrodynamics
models is devised. The Dirac type quantization procedure, based on the
canonical Hamiltonian formulation, is developed for some alternative
electrodynamics models. Within an approach developed a possibility of the
combined description both of electrodynamics and gravity is analyzed.

\section{Introduction}

\setcounter{page}{1} The classical relativistic electrodynamics of a freely
moving charged point particle in the Minkovski space-time $\mathbb{M}^{4}:=%
\mathbb{R}^{3}\mathbb{\times R}$ is, as well known, based \cite{LL,Fe1,Pa,Th}
\ on the Lagrangian formalism assigning to it the following Lagrangian
function
\begin{equation}
\mathcal{L}:=-m_{0}(1-u^{2})^{1/2},  \label{1.1}
\end{equation}%
where $m_{0}\in \mathbb{R}$ is the so-called particle rest mass and $u\in
\mathbb{E}^{3}$ is its spatial velocity in the Euclidean space $\mathbb{E}%
^{3},$ expressed here and throughout further in the light speed units (that
is the light speed equals $c=1).$ The least action Fermat principle in the
form
\begin{equation}
\delta S=0,\text{ \ \ }S:=-\int_{t_{1}}^{t_{2}}m_{0}(1-u^{2})^{1/2}dt
\label{1.2}
\end{equation}%
for any fixed \ temporal interval $[t_{1},t_{2}]\subset \mathbb{R}$ gives
rise to the well known relativistic relationships for the mass of the
particle
\begin{equation}
m=m_{0}(1-u^{2})^{-1/2},  \label{1.3}
\end{equation}%
the momentum of the particle

\begin{equation}
p:=mu=m_{0}u(1-u^{2})^{-1/2}  \label{1.4}
\end{equation}%
and the energy of the particle%
\begin{equation}
E_{0}=m=m_{0}(1-u^{2})^{-1/2}.  \label{1.5}
\end{equation}%
The origin of Lagrangian (\ref{1.1}), owing to the reasonings from \cite%
{LL,Pa}, can be extracted from the action expression%
\begin{equation}
S:=-\underset{t_{1}}{\overset{t_{2}}{\int }}m_{0}(1-u^{2})^{1/2}dt=-\underset%
{\tau _{1}}{\overset{\tau _{2}}{\int }}m_{0}d\tau ,  \label{1.6}
\end{equation}%
on the suitable temporal interval $[\tau _{1,}\tau _{2}]\subset \mathbb{R}%
\mathbf{,}$ where,\textbf{\ }by definition, $d\tau :=dt(1-u^{2})$$^{1/2}$
and $\tau \in \mathbb{R}$ is, so-called, proper temporal parameter assigned
to a freely moving particle with respect to the \ "rest" reference system. \
The action (\ref{1.6}) looks from the dynamical point of view sightly
controversial, since it is physically defined with respect to the \ "rest"
reference system, giving rise to the constant action $S=-m_{0}(\tau
_{2}-\tau _{1}),$ as limits of integrations $\tau _{1}<\tau _{2}\in \mathbb{R%
}$ were taken to be fixed from the very beginning. Moreover, considering
this particle as charged with a charge $q\in \mathbb{R}$ and moving in the
Minkovski space-time $\mathbb{M}^{4}$ under action of an electromagnetic
field $(\varphi ,A)\in \mathbb{R}\times \mathbb{E}^{3},$ the corresponding
classical (relativistic) action functional is chosen\ (see \cite%
{LL,Fe1,Pa,Th}) \ as follows:
\begin{equation}
S:=\underset{\tau _{1}}{\overset{\tau _{2}}{\int }}[-m_{0}d\tau +q<A,\dot{r}%
>d\tau -q\varphi (1-u^{2})^{-1/2}d\tau ],  \label{1.7}
\end{equation}%
with respect to the so-called \ "rest" reference system, parameterized by
the Euclidean space-time variables $(r,\tau )\in \mathbb{E}^{4},$ where $%
<\cdot ,\cdot >$ \ is the standard scalar product in the related Euclidean
subspace $\mathbb{E}^{3}$ and there is denoted $\dot{r}:=dr/d\tau $ in
contrast to the definition $u:=dr/dt.$ The action (\ref{1.7}) can be
rewritten, with respect to the moving with velocity\ vector $u\in \mathbb{R}%
^{3}$ reference system, as
\begin{equation}
S=\underset{t_{1}}{\overset{t_{2}}{\int }}\mathcal{L}dt,\text{ \ }\mathcal{L}%
:=-m_{0}(1-u^{2})^{1/2}+q<A,u>-q\varphi ,  \label{1.8}
\end{equation}%
\ on the suitable temporal interval \bigskip $\lbrack t_{1},t_{2}]\subset
\mathbb{R},$ giving rise to the following \cite{LL,Fe1,Pa,Th} \ dynamical
expressions
\begin{equation}
P=p+qA,\text{ \ \ \ \ }p=mu,  \label{1.9}
\end{equation}%
for the particle momentum and
\begin{equation}
E_{0}=[m_{0}^{2}+(P-qA)^{2}]^{1/2}+q\varphi  \label{1.10}
\end{equation}%
for the particle energy, where, by definition, $P\in \mathbb{E}^{3}$ means
the common momentum of the particle and the ambient electromagnetic field at
a space-time point $(r,t)\in \mathbb{M}^{4}.$

The obtained expression (\ref{1.10}) for the particle energy $E_{0}$ also
looks slightly controversial, since the potential energy $q\varphi ,$
entering additively, has no impact into the particle mass $%
m=m_{0}(1-u^{2})^{-1/2}.$ As it was already mentioned \cite{Br} by L.
Brillouin, the fact that the potential energy has no impact to the particle
mass says us that "... any possibility of existing the particle mass related
with an external potential energy, is completely excluded". This and some
other special relativity theory and electrodynamics problems, as is well
known, stimulated many other prominent physicists\ of the past \cite%
{Br,Fe,We,Pa,BD} and the present \cite{Me,Me1,Lo,Lo1,Lo2,Lo3,B-B,Re,Ne} to
make significant efforts aiming to develop alternative relativity theories
based on completely different space-time and matter structure principles.

There also is another controversial inference from the action expression (%
\ref{1.8}). As one can easily show \cite{LL}, the corresponding dynamical
equation \ for the Lorentz force is given as follows:
\begin{equation}
dp/dt=F:=qE+qu\times B,  \label{1.11}
\end{equation}%
where the operation $"\times "$ denotes the standard vector product and we
put, by definition,
\begin{equation}
E:=-\partial A/\partial t-\nabla \varphi   \label{1.12}
\end{equation}%
for the related electric field and
\begin{equation}
B:=\nabla \times A  \label{1.13}
\end{equation}%
for the related magnetic field, acting on the charged point particle $q;$
the operation $"\nabla "$ is the standard gradient. The obtained expression (%
\ref{1.11}) means, in particular, that the Lorentz force $F$\ depends
linearly on the particle velocity vector $u\in \mathbb{E}^{3},$ giving rise
to its strong dependence on the reference system with respect to which the
charged particle $q$ moves. Namely, the attempts to reconcile this and some
related controversies \cite{Br,Fe,Re,Kl} forced A. Einstein to devise his
special relativity theory and proceed further to creating his general
relativity theory trying to explain the gravity by means of a geometrization
of space-time and matter in the Universe. Here we must mention that the
classical Lagrangian function $\mathcal{L}$ in (\ref{1.8}) is written by
means of the mixed combinations of terms expressed by means of both the
Euclidean "rest" reference system variables $(r,\tau )\in \mathbb{E}^{4}$
and an arbitrarily chosen reference system variables $(r,t)\in \mathbb{M}%
^{4}.$

These problems were recently analyzed from a completely another
"no-geometry" \ point of view in \cite{BPT,BPT1,Re}, where new dynamical
equations were derived, being free of controversy mentioned above. Moreover,
the devised approach allowed to avoid the introduction of the well known
Lorentz transformations of the space-time reference systems with respect to
which the action functional (\ref{1.8}) is invariant. Below we will
reanalyze the results obtained in \cite{BPT,BPT1} from the classical
Lagrangian and Hamiltonian formalisms, what will shed a new light on the
related physical backgrounds of the vacuum field theory approach to common
studying electromagnetic and gravitational effects.

\section{The vacuum field theory electrodynamics equations: Lagrangian
analysis}

Within the vacuum field theory approach to common describing the
electromagnetism and the gravity, devised in \cite{BPT,BPT1}, the main
vacuum potential field $\bar{W}:\mathbb{M}^{4}\mathbb{\rightarrow R},$
related with a charged point particle $q,$ satisfies in the case of the
rested external charged point objects the following dynamical equation
\begin{equation}
\frac{d}{dt}(-\bar{W}u)=-\nabla \bar{W},  \label{2.1}
\end{equation}%
where, as above, $u:=dr/dt$ is the particle velocity with respect to some
reference system.

To analyze the dynamical equation (\ref{2.1}) from the Lagrangian point of
view we will write the corresponding action functional as
\begin{equation}
S:=-\underset{t_{1}}{\overset{t_{2}}{\int }}\bar{W}dt=-\underset{\tau _{1}}{%
\overset{\tau _{2}}{\int }}\bar{W}(1+\dot{r}^{2})^{1/2}\text{ }d\tau ,
\label{2.2}
\end{equation}%
expressed with respect to the "rest" reference system. Having fixed proper
temporal parameters $\tau _{1}<\tau _{2}\in \mathbb{R},$ from the least
action condition $\delta S=0$ one finds easily that
\begin{eqnarray}
p &:&=\partial \mathcal{L}/\partial \dot{r}=-\bar{W}\dot{r}(1+\dot{r}%
^{2})^{-1/2}=-\bar{W}u,  \label{2.3} \\
\dot{p} &:&=dp/d\tau =\partial \mathcal{L}/\partial r=-\nabla \bar{W}(1+\dot{%
r}^{2})^{1/2},  \notag
\end{eqnarray}%
where, owing to (\ref{2.2}), the corresponding Lagrangian function
\begin{equation}
\mathcal{L}:=-\bar{W}(1+\dot{r}^{2})^{1/2}.  \label{2.4}
\end{equation}%
Recalling now the definition of the particle mass
\begin{equation}
m:=-\bar{W}  \label{2.5}
\end{equation}%
and \ the relationships%
\begin{equation}
d\tau =dt(1-u^{2})^{1/2},\text{ }\dot{r}d\tau =udt,  \label{2.6}
\end{equation}%
from (\ref{2.3}) we easily obtain exactly the dynamical equation (\ref{2.1}).

Proceed now to the case when our charged point particle $q$ moves in the
space time with velocity vector $u\in \mathbb{E}^{3}$ and interacts with
another external charged point particle, moving with velocity vector $v\in
\mathbb{E}^{3}$ subject to some common reference system $\mathcal{K}.$ As
was shown in \cite{BPT,BPT1}, the corresponding dynamical equation on the
vacuum potential field function $\bar{W}:\mathbb{M}^{4}\mathbb{\rightarrow R}
$ is given as
\begin{equation}
\frac{d}{dt}[-\bar{W}(u-v)]=-\nabla \bar{W}.  \label{2.7}
\end{equation}%
As the external charged particle moves in the space-time, it generates the
related magnetic field $B:=\nabla \times A,$ whose magnetic vector potential
$A:\mathbb{M}^{4}\mathbb{\rightarrow E}^{3}$ is defined, owing to the
results of \cite{BPT,BPT1,Re}, as
\begin{equation}
qA:=\bar{W}v.  \label{2.8}
\end{equation}%
Since, owing to (\ref{2.3}), the particle momentum $p=-\bar{W}u,$ equation (%
\ref{2.7}) can be equivalently rewritten as
\begin{equation}
\frac{d}{dt}(p+qA)=-\nabla \bar{W}.  \label{2.9}
\end{equation}%
To represent the dynamical equation (\ref{2.9}) within the classical
Lagrangian formalism, we start from the following action functional
naturally generalizing functional (\ref{2.2}):%
\begin{equation}
S:=-\underset{\tau _{1}}{\overset{\tau _{2}}{\int }}\bar{W}(1+|\dot{r}-\dot{%
\xi}^{2}|)^{1/2}\text{ }d\tau ,  \label{2.10}
\end{equation}%
where we denoted by $\dot{\xi}=vdt/d\tau ,$ $d\dot{\tau}%
=dt(1-|u-v|^{2})^{1/2},$ which take into account the \ relative velocity of
our charged point particle $q$ with respect to the reference system $%
\mathcal{K}^{\prime },$ moving with velocity vector $v\in \mathbb{E}^{3}$
subject to the reference system $\mathcal{K}.$ In this case, evidently, our
charged point particle $q$ moves with velocity vector $u-v\in \mathbb{E}^{3}$
subject to the reference system $\mathcal{K}^{\prime },$ and the external
charged particle is, respectively, in rest.

Compute now the least action variational condition $\delta S=0$, taking into
account that, owing to (\ref{2.10}), the corresponding Lagrangian function
is given as
\begin{equation}
\mathcal{L}:=-\bar{W}(1+|\dot{r}-\dot{\xi}^{2}|)^{1/2}.  \label{2.11}
\end{equation}%
Thereby, the common particles momentum
\begin{eqnarray}
P &:&=\partial \mathcal{L}/\partial \dot{r}=-\bar{W}(\dot{r}-\dot{\xi})(1+|%
\dot{r}-\dot{\xi}^{2}|^{2})^{-1/2}=  \label{2.12} \\
&=&-\bar{W}\dot{r}(1+|\dot{r}-\dot{\xi}^{2}|^{2})^{-1/2}+\bar{W}\dot{\xi}(1+|%
\dot{r}-\dot{\xi}^{2}|^{2})^{-1/2}=  \notag \\
&=&mu+qA:=p+qA  \notag
\end{eqnarray}%
and the dynamical equation is given as
\begin{equation}
\frac{d}{d\tau }(p+qA)=-\nabla \bar{W}(1+|\dot{r}-\dot{\xi}^{2}|)^{1/2}.
\label{2.13}
\end{equation}%
Taking into account that $d\tau =dt(1-\left\vert u-v\right\vert ^{2})^{1/2}$
and $(1+|\dot{r}-\dot{\xi}^{2}|^{2})^{1/2},$ we obtain finally from (\ref%
{2.13}) exactly the dynamical equation (\ref{2.9}).

\begin{remark}
\textit{{It is easy to observe that the action functional (\ref{2.10}) is
written taking into account the classical Galilean transformations of
reference systems. If we now consider the action functional (\ref{2.2}) for
a charged point particle and take into account its interaction with an
external magnetic field, generated by the vector potential $A:$ $\mathbb{M}%
^{4}\mathbb{\rightarrow E}^{3},$ it can be naturally generalized as
\begin{equation}
S:=\underset{t_{1}}{\overset{t_{2}}{\int }}(-\bar{W}dt+q<A,dr>)=\underset{%
\tau _{1}}{\overset{\tau _{2}}{\int }}[-\bar{W}(1+\dot{r}^{2})^{1/2}\text{ }%
+q<A,\dot{r}>]d\tau ,  \label{2.14}
\end{equation}%
where we accepted here that $d\tau =dt(1-u^{2})^{1/2}.$} }
\end{remark}

Thus, the corresponding common particle-field momentum looks as follows:
\begin{eqnarray}
P &:&=\partial \mathcal{L}/\partial \dot{r}=-\bar{W}\dot{r}(1+\dot{r}%
^{2})^{-1/2}+qA=  \label{2.15} \\
&=&mu+qA:=p+qA,  \notag
\end{eqnarray}%
satisfying the equation%
\begin{eqnarray}
dP/d\tau &:&=\partial \mathcal{L}/\partial r=-\nabla \bar{W}(1+\dot{r}%
^{2})^{1/2}\text{ }+q\nabla <A,\dot{r}>=  \label{2.16} \\
&=&-\nabla \bar{W}(1-u^{2})^{-1/2}+q\nabla <A,u>(1-u^{2})^{-1/2},  \notag
\end{eqnarray}%
where
\begin{equation}
\mathcal{L}:=-\bar{W}(1+\dot{r}^{2})^{1/2}\text{ }+q<A,\dot{r}>
\label{2.16a}
\end{equation}%
is the corresponding Lagrangian function. Taking now into account \ that $%
d\tau =dt(1-u^{2})^{1/2},$ one easily finds from (\ref{2.16}) that
\begin{equation}
dP/dt=-\nabla \bar{W}+q\nabla <A,u>.  \label{2.17}
\end{equation}%
Upon substituting (\ref{2.15}) into (\ref{2.17}) and making use of the well
known \cite{LL} identity
\begin{equation}
\nabla <a,b>=<a,\nabla >b+<b,\nabla >a+b\times (\nabla \times a)+a\times
(\nabla \times b),  \label{2.18}
\end{equation}%
where $a,b\in \mathbb{E}^{3}$ are arbitrary vector functions, we obtain
finally the classical expression for the Lorentz force $F,$ acting on the
moving charged point particle $q:$%
\begin{equation}
dp/dt:=F=qE+qu\times B,  \label{2.19}
\end{equation}%
where, by definition,
\begin{equation}
E:=-\nabla \bar{W}q^{-1}-\partial A/\partial t  \label{2.20}
\end{equation}%
is the corresponding electric field and
\begin{equation}
B:=\nabla \times A  \label{2,21}
\end{equation}%
is the corresponding magnetic field.

Concerning the previously obtained dynamical equation (\ref{2.13}) we can
easily observe that it can be equivalently rewritten as follows:
\begin{equation}
dp/dt=(-\nabla \bar{W}-qdA/dt+q\nabla <A,u>)-q\nabla <A,u>.  \label{2.22}
\end{equation}%
The latter, owing to (\ref{2.17}) and (\ref{2.19}), takes finally the
following Lorentz type force in the form%
\begin{equation}
dp/dt=qE+qu\times B-q\nabla <A,u>,  \label{2.23}
\end{equation}%
before found in \cite{BPT,BPT1,Re}.

Expressions (\ref{2.19}) and (\ref{2.23}) are equal to each other up to the
gradient term $F_{c}:=-q\nabla <A,u>$, which allows to reconcile the Lorentz
forces acting on a charged moving particle $q$ with respect to different
reference systems. This fact is important for our vacuum field theory
approach since it needs to use no special geometry and makes it possible to
analyze both electromagnetic and gravitational fields simultaneously, based
on a new definition of the dynamical mass by means of expression (\ref{2.5}).

\section{The vacuum field theory electrodynamics equations: Hamiltonian
analysis}

It is well know \cite{Ar,Th,AM,HK,PM} that any Lagrangian theory allows the
equivalent canonical Hamiltonian representation via the classical Legendrian
transformation. As we have already formulated above our vacuum field theory
of a moving charged particle $q$ in the Lagrangian form, we proceed now to
its Hamiltonian analysis making use of the action functionals (\ref{2.2}), (%
\ref{2.11}) and (\ref{2.14}).

Take, first, the Lagrangian function (\ref{2.4}) and the momentum expression
(\ref{2.3}) for defining the corresponding Hamiltonian function%
\begin{eqnarray}
H &:&=<p,\dot{r}>-\mathcal{L}=  \notag \\
&=&-<p,p>\bar{W}^{-1}(1-p^{2}/\bar{W}^{2})^{-1/2}+\bar{W}(1-p^{2}/\bar{W}%
^{2})^{-1/2}=  \notag \\
&=&-p^{2}\bar{W}(1-p^{2}/\bar{W}^{2})^{-1/2}+\bar{W}^{2}\bar{W}(1-p^{2}/\bar{%
W}^{2})^{-1/2}=  \label{3.1} \\
&=&(\bar{W}^{2}-p^{2})(\bar{W}^{2}-p^{2})^{-1/2}=(\bar{W}^{2}-p^{2})^{1/2}.
\notag
\end{eqnarray}%
As a result, we easily obtain \cite{AM,Ar,Th,PM} \ that the Hamiltonian
function (\ref{3.1}) is a conservation law of the dynamical field equation (%
\ref{2.1}), that is for all $\tau ,t\in \mathbb{R}$%
\begin{equation}
dH/dt=0=dH/d\tau ,  \label{3.2}
\end{equation}%
which naturally \ allows \ to interpret it as the energy expression. Thus,
we can write that the particle energy
\begin{equation}
E=(\bar{W}^{2}-p^{2})^{1/2}.  \label{3.3}
\end{equation}%
The corresponding Hamiltonian equations, equivalent to the vacuum field
equation (\ref{2.1}), look as follows:%
\begin{eqnarray}
\dot{r} &:&=\partial H/\partial p=-p(\bar{W}^{2}-p^{2})^{-1/2}  \label{3.4}
\\
\dot{p} &:&=-\partial H/\partial r=-\bar{W}\nabla \bar{W}(\bar{W}%
^{2}-p^{2})^{-1/2}.  \notag
\end{eqnarray}

Based now on the Lagrangian expression (\ref{2.1}) one can construct, the
same way as above, the Hamiltonian function for the dynamical field equation
(\ref{2.9}), describing the motion of charged particle $q$ in external
electromagnetic field in the canonical Hamiltonian form:%
\begin{equation}
\dot{r}:=\partial H/\partial p,\text{ \ \ \ \ \ }\dot{p}:=-\partial
H/\partial r,  \label{3.5a}
\end{equation}%
where%
\begin{eqnarray}
H &:&=<P,\dot{r}>-\mathcal{L}=  \notag \\
&=&<P,\dot{\xi}-P\bar{W}^{-1}(1-p^{2}/\bar{W}^{2})^{-1/2}>+\bar{W}[\bar{W}%
^{2}(\bar{W}^{2}-p^{2})^{-1}]^{1/2}=  \notag \\
&=&<P,\dot{\xi}>-P^{2}\bar{W}^{-1}(1-P^{2}/\bar{W}^{2})^{-1/2}+\bar{W}%
(1-p^{2}/\bar{W}^{2})^{-1/2}=  \notag \\
&=&(\bar{W}^{2}-p^{2})(\bar{W}^{2}-p^{2})^{-1/2}+<P,\dot{\xi}>=  \label{3.5}
\\
&=&(\bar{W}^{2}-P^{2})^{1/2}+q<A,P>(\bar{W}^{2}-P^{2})^{-1/2}.  \notag
\end{eqnarray}%
Here we took into account that, owing to definitions (\ref{2.8}) and (\ref%
{2.12}),%
\begin{eqnarray}
qA &:&=\bar{W}v=\bar{W}d\xi /dt=  \label{3.6} \\
&=&\bar{W}\frac{d\xi }{d\tau }\cdot \frac{d\tau }{dt}=\bar{W}\dot{\xi}%
(1-\left\vert u-v\right\vert ^{2})^{1/2}=  \notag \\
&=&\bar{W}\dot{\xi}(1+|\dot{r}-\dot{\xi}|^{2})^{-1/2}=  \notag \\
&=&\bar{W}\dot{\xi}(\bar{W}^{2}-P^{2})^{1/2}\bar{W}^{-1}=\dot{\xi}(\bar{W}%
^{2}-P^{2})^{1/2},  \notag
\end{eqnarray}%
or%
\begin{equation}
\dot{\xi}=qA(\bar{W}^{2}-P^{2})^{-1/2},  \label{3.7}
\end{equation}%
where $A:\mathbb{M}^{4}\mathbb{\rightarrow R}^{3}$ is the related magnetic
vector potential, generated by the moving external charged particle.

Thereby we can state that the Hamiltonian function (\ref{3.5}) satisfies the
energy conservation condition%
\begin{equation}
dH/dt=0=dH/d\tau ,  \label{3.8}
\end{equation}%
for all $\tau ,t\in \mathbb{R},$ that is the corresponding energy expression%
\begin{equation}
E=(\bar{W}^{2}-P^{2})^{1/2}+q<A,P>(\bar{W}^{2}-P^{2})^{-1/2}  \label{3.9}
\end{equation}%
holds. The result (\ref{3.9}) essentially differs from that obtained in \cite%
{LL}, which makes use of the well known Einsteinian Lagrangian for a moving
charged point particle $q$ in external electromagnetic field.

To make this difference more clear, we will analyze below the Lorentz force (%
\ref{2.19}) from the Hamiltonian point of view based on the Lagrangian
function (\ref{2.16a}). Thus, we obtain that the corresponding Hamiltonian
function%
\begin{eqnarray}
H &:&=<P,\dot{r}>-\mathcal{L}=<P,\dot{r}>+\bar{W}(1+\dot{r}^{2})^{1/2}-q<A,%
\dot{r}>=  \label{3.10} \\
&=&<P-qA,\dot{r}>+\bar{W}(1+\dot{r}^{2})^{1/2}=  \notag \\
&=&-<p,p>\bar{W}^{-1}(1-p^{2}/\bar{W}^{2})^{-1/2}+\bar{W}(1-p^{2}/\bar{W}%
^{2})^{-1/2}=  \notag \\
&=&(\bar{W}^{2}-p^{2})(\bar{W}^{2}-p^{2})^{-1/2}=(\bar{W}^{2}-p^{2})^{1/2}.
\notag
\end{eqnarray}%
Since $p=P-qA,$ expression (\ref{3.10}) takes the final form as
\begin{equation}
H=[\bar{W}^{2}-(P-qA)^{2}]^{1/2},  \label{3.11}
\end{equation}%
being conservative with respect to the evolution equations (\ref{2.15}) and (%
\ref{2.16}), that is
\begin{equation}
dH/dt=0=dH/d\tau  \label{3.11a}
\end{equation}%
for all $\tau ,t\in \mathbb{R}.$ The latter are simultaneously equivalent to
the following Hamiltonian system:%
\begin{eqnarray}
\dot{r} &=&\partial H/\partial P=(qA-P)[\bar{W}^{2}-(P-qA)^{2}]^{1/2},
\label{3.12} \\
\dot{P} &=&-\partial H/\partial r=(-\bar{W}\nabla \bar{W}-\nabla
<qA,(P-qA)>)[\bar{W}^{2}-(P-qA)^{2}]^{1/2},  \notag
\end{eqnarray}%
that can be easily checked by direct calculations. Really, the first equation%
\begin{eqnarray}
\dot{r} &=&(qA-p)[\bar{W}^{2}-(p-qA)^{2}]^{-1/2}=-p(\bar{W}%
^{2}-p^{2})^{-1/2}=  \label{3.13} \\
&=&-mu(\bar{W}^{2}-p^{2})^{-1/2}=\bar{W}u(\bar{W}%
^{2}-p^{2})^{-1/2}=u(1-u^{2})^{-1/2},  \notag
\end{eqnarray}%
holds, owing to the condition $d\tau =dt(1-u^{2})^{1/2}$ and definitions $%
p:=mu,$ $m=-\bar{W},$ postulated from the very beginning. Similarly we
obtain that
\begin{eqnarray}
\dot{P} &=&-\nabla \bar{W}(1-p^{2}/\bar{W}^{2})^{-1/2}+\nabla <qA,\bar{W}%
u>(1-p^{2}/\bar{W}^{2})^{-1/2}=  \label{3.14} \\
&=&-\nabla \bar{W}(1-u^{2})^{-1/2}+\nabla <qA,u>(1-u^{2})^{-1/2},  \notag
\end{eqnarray}%
exactly coinciding with equation (\ref{2.17}) subject to the evolution
parameter $t\in \mathbb{R}.$

\section{The quantization of electrodynamics models within the vacuum field
theory no-geometry approach}

\subsection{The problem setting}

In our recent works \cite{BPT,BPT1} there was devised a new regular
no-geometry approach to deriving from the first principles the
electrodynamics of a moving charged point particle $q$ in external
electromagnetic field. This approach has, in part, to reconcile the existing
mass-energy controversy \cite{Br} within the classical relativistic
electrodynamics. Based on the vacuum field theory approach proposed in \cite%
{BPT,BPT1,Re} we reanalyzed this problem in sections above both from
Lagrangian and Hamiltonian points of view having derived crucial expressions
for the corresponding energy functions and Lorentz type forces, acting on
moving charge point particle $q.$

Since all of our electrodynamics models were represented here in the
canonical Hamiltonian form, they are suitable for applying to them the Dirac
type quantization procedure \cite{Di,BS,BS1} and regular obtaining the
related Schr\"{o}dinger type evolution equations. Namely, to this problem
there is devoted this Section.

\subsection{Free point particle electrodynamics model and its quantization}

The charged point particle electrodynamics models, discussed in detail in
Sections 2 and 3, were also considered in \cite{BPT1} from the dynamical
point of view, where an attempt of application the quantization Dirac type
procedure to the corresponding conserved energy expressions was done.
Nevertheless, within the canonical point of view, \ the true quantization
procedure should be based on the suitable canonical Hamiltonian formulation
of the models, which in the case under consideration looks as (\ref{3.4}), (%
\ref{3.5a}) and (\ref{3.12}).

In particular, consider a free charged \ point particle electrodynamics
model, governed (\ref{3.4}) by the following Hamiltonian equations:%
\begin{eqnarray}
dr/d\tau &:&=\partial H/\partial p=-p(\bar{W}^{2}-p^{2})^{-1/2},  \label{5.1}
\\
dp/d\tau &:&=-\partial H/\partial r=-\bar{W}\nabla \bar{W}(\bar{W}%
^{2}-p^{2})^{-1/2},  \notag
\end{eqnarray}%
where we denoted, as before, by $\bar{W}:\mathbb{M}^{4}\rightarrow \mathbb{R}
$ the corresponding vacuum field potential, characterizing medium field
structure, by $(r,p)\in \mathbb{E}^{3}\times \mathbb{E}^{3}$ the standard
canonical coordinate-momentum variables, by $\tau \in \mathbb{R}$ the proper
"rest" reference system $\mathcal{K}_{r}$ time parameter, related with our
moving particle, and by $H:\mathbb{E}^{3}\times \mathbb{E}^{3}\rightarrow
\mathbb{R}$ the Hamiltonian function%
\begin{equation}
H:=(\bar{W}^{2}-p^{2})^{1/2},  \label{5.2}
\end{equation}%
expressed here and throughout further in the light speed units. The "rest"
reference system $\mathcal{K}_{r},$ parameterized by variables $(r,\tau )\in
\mathbb{E}^{4},$ is related with any other reference system $\mathcal{K}$
subject to which our charged point particle $q$ moves with velocity vector $%
u\in \mathbb{E}^{3},$ and which is parameterized by variables $(r,t)\in
\mathbb{M}^{4},$ via the following Euclidean infinitesimal relationship:%
\begin{equation}
dt^{2}=d\tau ^{2}+dr^{2},  \label{5.3}
\end{equation}%
which is equivalent to the Minkovskian infinitesimal relationship%
\begin{equation}
d\tau ^{2}=dt^{2}-dr^{2}.  \label{5.4}
\end{equation}%
The Hamiltonian function (\ref{5.2}) satisfies, evidently, the energy
conservation conditions
\begin{equation}
dH/dt=0=dH/d\tau  \label{5.5}
\end{equation}%
for all $t,\tau \in \mathbb{R}.$ This means that the energy value
\begin{equation}
E=(\bar{W}^{2}-p^{2})^{1/2}  \label{5.6}
\end{equation}%
can be treated by means of the Dirac type quantization scheme \cite{Di} to
obtain, as $\hbar \rightarrow 0,$ (or the light speed $c\rightarrow \infty $%
) the governing Schr\"{o}dinger type dynamical equation. To do this,
similarly to \cite{BPT,BPT1}, we need to make canonical operator
replacements $E\rightarrow \hat{E}:=-\frac{\hbar }{i}\frac{\partial }{%
\partial \tau },$ \ $p\rightarrow \hat{p}:=\frac{\hbar }{i}\nabla ,$ as $%
\hbar \rightarrow 0,$ in the following energy determining expression:%
\begin{equation}
E^{2}:=(\hat{E}\psi ,\hat{E}\psi )=(\psi ,\hat{E}^{2}\psi )=(\psi ,\hat{H}%
^{+}\hat{H}\psi ),  \label{5.7}
\end{equation}%
where, by definition, owing to (\ref{5.6}),
\begin{equation}
\hat{E}^{2}=\bar{W}^{2}-\hat{p}^{2}=\hat{H}^{+}\hat{H}  \label{5.8}
\end{equation}%
is a suitable operator factorization in the Hilbert space $\mathcal{H}:%
\mathcal{=}L_{2}(\mathbb{R}^{3};\mathbb{C})$ and $\psi \in \mathcal{H}$ is
the corresponding normalized quantum vector state. Since the following
elementary identity
\begin{equation}
\bar{W}^{2}-\hat{p}^{2}=\bar{W}(1-\bar{W}^{-1}\hat{p}^{2}\bar{W}%
^{-1})^{1/2}(1-\bar{W}^{-1}\hat{p}^{2}\bar{W}^{-1})^{1/2}\bar{W}  \label{5.9}
\end{equation}%
holds, we can put, by definition, following (\ref{5.8}) and (\ref{5.9}) that
the operator
\begin{equation}
\hat{H}:=(1-\bar{W}^{-1}\hat{p}^{2}\bar{W}^{-1})^{1/2}\bar{W}.  \label{5.10}
\end{equation}%
Having calculated the operator expression (\ref{2.10}) as $\hbar \rightarrow
0$ up to operator accuracy $O($ $\hbar ^{4}),$ \ we can easily obtain that
\begin{equation}
\hat{H}=\frac{\hat{p}^{2}}{2m(u)}+\bar{W}:=-\frac{\hbar ^{2}}{2m(u)}\nabla
^{2}+\bar{W},  \label{5.11}
\end{equation}%
where we took into account the dynamical mass definition $m(u):=-\bar{W}$
(in the light speed units). Thereby, based now on (\ref{5.7}) and (\ref{5.11}%
), we obtain up to operator accuracy $O($ $\hbar ^{4})$ the following Schr%
\"{o}dinger type evolution equation%
\begin{equation}
i\hbar \frac{\partial \psi }{\partial \tau }:=\hat{E}\psi =\hat{H}\psi =-%
\frac{\hbar ^{2}}{2m(u)}\nabla ^{2}\psi +\bar{W}\psi  \label{5.12}
\end{equation}%
with respect to the "rest" \ reference system $\mathcal{K}_{r}$ evolution
parameter $\tau \in \mathbb{R}.$ Concerning the related evolution parameter $%
t\in \mathbb{R},$ parameterizing a reference system $\mathcal{K},$ the
equation (\ref{5.12}) takes the following form:%
\begin{equation}
i\hbar \frac{\partial \psi }{\partial t}=-\frac{\hbar ^{2}m_{0}}{2m(u)^{2}}%
\nabla ^{2}\psi -m_{0}\psi .  \label{5.13}
\end{equation}%
Here we took into account that, owing to (\ref{5.6}), the classical mass
relationship
\begin{equation}
m(u)=m_{0}(1-u^{2})^{-1/2}  \label{5.14}
\end{equation}%
holds, where $m_{0}\in \mathbb{R}_{+}$ is the corresponding rest mass of our
point particle $q.$

The obtained linear \ \ Schr\"{o}dinger equation (\ref{5.13}) for the case $%
\hbar /c\rightarrow 0$ really coincides with that well-known \cite{LL,Di,Fe1}
from classical quantum mechanics.

\subsection{Classical charged point particle electrodynamics model and its
quantization}

We start here from the first vacuum field theory reformulation of the
classical charged point particle electrodynamics, considered in Section 3
and based on the conserved Hamiltonian function (\ref{3.11})
\begin{equation}
H:=[\bar{W}^{2}-(P-qA)^{2}]^{1/2},  \label{6.1}
\end{equation}%
where $q\in $ $\mathbb{R}$ is the particle charge and $(\bar{W},A)\in
\mathbb{R\times E}^{3}$ is the corresponding electromagnetic field
potentials and $P\in \mathbb{E}^{3}$ is the common particle-field momentum,
defined as
\begin{equation}
P:=p+qA,\text{ \ \ \ \ }p:=mu,  \label{6.2}
\end{equation}%
and satisfying the well known classical Lorentz force equation. Here $m:=-%
\bar{W}$ is the observable dynamical mass, related with our charged
particle, $u\in \mathbb{E}^{3}$ is its velocity vector with respect to a
chosen reference system $\mathcal{K},$ being all expressed here, as before,
in the light speed units.

As our electrodynamics, based on (\ref{6.1}), is canonically Hamiltonian,
the Dirac type quantization scheme
\begin{equation}
P\rightarrow \hat{P}:=\frac{\hbar }{i}\nabla ,\text{ \ \ \ \ \ \ }%
E\rightarrow \hat{E}:=-\frac{\hbar }{i}\frac{\partial }{\partial \tau }
\label{6.3}
\end{equation}%
should be applied to the energy expression
\begin{equation}
E:=[\bar{W}^{2}-(P-qA)^{2}]^{1/2},  \label{6.4}
\end{equation}%
following from the conservation conditions
\begin{equation}
dH/dt=0=dH/d\tau ,  \label{6.5}
\end{equation}%
satisfied for all $\tau ,t\in \mathbb{R}.$

Doing now the same way as above, we can factorize the operator $\hat{E}^{2}$
as follows:%
\begin{equation*}
\begin{array}{c}
\bar{W}^{2}-(\hat{P}-qA)^{2}=\bar{W}[1-\bar{W}^{-1}(\hat{P}-qA)^{2}\bar{W}%
^{-1}]^{1/2}\times \\
\times \lbrack 1-\bar{W}^{-1}(\hat{P}-qA)^{2}\bar{W}^{-1}]^{1/2}\bar{W}:=%
\hat{H}^{+}\hat{H},%
\end{array}%
\end{equation*}%
where, by definition, (here as $\hbar /c\rightarrow 0,$ $\hbar c=const$)%
\begin{equation}
\hat{H}:=\frac{1}{2m(u)}(\frac{\hbar }{i}\nabla -qA)^{2}+\bar{W}  \label{6.7}
\end{equation}%
up to operator accuracy $O(\hbar ^{4}).$ Thereby, the related Schr\"{o}%
dinger type evolution equation in the Hilbert space $\mathcal{H=}L_{2}(%
\mathbb{R}^{3};\mathbb{C})$ looks as
\begin{equation}
\hbar \frac{\partial \psi }{\partial \tau }:=\hat{E}\psi =\hat{H}\psi =\frac{%
1}{2m(u)}(\frac{\hbar }{i}\nabla -qA)^{2}\psi +\bar{W}\psi  \label{6.8}
\end{equation}%
with respect to the rest reference system $\mathcal{K}_{r}$ evolution
parameter $\tau \in \mathbb{R}.$ The corresponding Schr\"{o}dinger type
evolution equation with respect to the evolution parameter $t\in \mathbb{R}$
looks, respectively, as
\begin{equation}
\hbar \frac{\partial \psi }{\partial t}=\frac{m_{0}}{2m(u)^{2}}(\frac{\hbar
}{i}\nabla -qA)^{2}\psi -m_{0}\psi .  \label{6.9}
\end{equation}%
The Schr\"{o}dinger type evolution equation (\ref{6.8}) ( as $\hbar
/c\rightarrow 0,$ $\hbar c=const$) completely coincides \cite{LL1,Di} with
that well known from the classical quantum mechanics.

\subsection{Modified charged point particle electrodynamics model and its
quantization}

Coinsider now, within the canonical point of view, \ the true quantization
procedure the electrodynamics model, which looks as (\ref{2.13}) and whose
Hamiltonian function (\ref{3.5}) is
\begin{equation}
H:=(\bar{W}^{2}-P^{2})^{1/2}+q<A,P>(\bar{W}^{2}-P^{2})^{-1/2}.  \label{7.1}
\end{equation}%
This means that the energy function
\begin{equation}
E:=(\bar{W}^{2}-P^{2})^{1/2}+q<A,P>(\bar{W}^{2}-P^{2})^{-1/2},  \label{7.2}
\end{equation}%
where, as before,
\begin{equation}
P:=p+qA,\text{ \ \ \ \ }p:=mu,\text{ \ }m:=-\dot{W},  \label{7.3}
\end{equation}%
is a conserved quantity for (\ref{2.13}), which we will canonically quantize
via the Dirac procedure (\ref{6.3}). To make this, let us consider the
quantum condition%
\begin{equation}
E^{2}:=(\hat{E}\psi ,\hat{E}\psi )=(\psi ,\hat{E}^{2}\psi ),\text{ \ \ \ \ \
}(\psi ,\psi ):=1,  \label{7.4}
\end{equation}%
where, by definition, $\hat{E}:=-\frac{\hbar }{i}\frac{\partial }{\partial t}
$ and $\psi \in \mathcal{H=}L_{2}(\mathbb{R}^{3};\mathbb{C})$ is a suitable
normalized quantum state vector. Making now use of the energy function (\ref%
{7.2}), one can easily obtain that
\begin{equation}
E^{2}=\bar{W}^{2}-(P-qA)^{2}+q^{2}[<A,A>+<A,P>(\bar{W}^{2}-P^{2})^{-1}<P,A>],
\label{7.5}
\end{equation}%
which upon the canonical Dirac type quantization $P\rightarrow \hat{P}:=%
\frac{\hbar }{i}\nabla $ transforms into the symmetrized operator expression
\begin{equation}
\hat{E}^{2}=\bar{W}^{2}-(\hat{P}-qA)^{2}+q^{2}<A,A>+q^{2}<A,P>(\bar{W}%
^{2}-P^{2})^{-1}<P,A>.  \label{7.6}
\end{equation}%
Having factorized operator (\ref{7.6})\ in the form $\hat{E}^{2}=\hat{H}^{+}%
\hat{H},$ we obtain that up to operator accuracy $O(\hbar ^{4})$ (as $\hbar
/c\rightarrow 0,$ $\hbar c=const$)
\begin{eqnarray}
\hat{H} &:&=\frac{1}{2m(u)}(\frac{\hbar }{i}\nabla -qA)^{2}-  \label{7.7} \\
-\frac{q^{2}}{2m(u)} &<&A,A>-\frac{q^{2}}{2m^{3}(u)}<A,P><P,A>,  \notag
\end{eqnarray}%
where we put, as before, $m(u)=-\bar{W}$ in the light speed units. Thus,
owing to (\ref{7.4}) and (\ref{7.7}), the resulting Schr\"{o}dinger
evolution equation takes the form
\begin{eqnarray}
i\hbar \frac{\partial \psi }{\partial \tau } &:&=\hat{H}\psi =\frac{1}{2m(u)}%
(\frac{\hbar }{i}\nabla -qA)^{2}\psi -  \label{7.8} \\
-\frac{q^{2}}{2m(u)} &<&A,A>\psi -\frac{q^{2}}{2m^{3}(u)}<A,P><P,A>\psi
\notag
\end{eqnarray}%
with respect to the "rest" reference system proper evolution parameter $\tau
\in \mathbb{R}.$ Similarly one obtains also the related Schr\"{o}dinger type
evolution equation with respect to the time parameter $t\in \mathbb{R}$ on
which we will not here stop. The result (\ref{7.8}) essentially differs the
corresponding classical Schr\"{o}dinger evolution equation (\ref{6.8}) that,
thereby, forces us to reanalyze more thoroughly the main physically
motivated principles, put into the backgrounds of classical electrodynamical
models, described by the Hamiltonian functions (\ref{6.1}) and (\ref{7.1})
giving rise to different Lorentz type force expressions. This analysis we
plan to do in detail in a next work under preparation.

\section{ Conclusion}

Thereby, we can claim that all of dynamical field equations discussed above
are canonical Hamiltonian systems with respect to the corresponding proper
"rest" reference systems, parameterized by suitable time parameters $\tau
\in \mathbb{R}.$ Owing to the passing to the basic reference system $%
\mathcal{K}$ with the time parameter $t\in \mathbb{R}$ the related
Hamiltonian structure is naturally lost, giving rise to a new interpretation
of the real particle motion as such having the absolute sense only with
respect to the proper "rest" \ reference system and being completely
relative with respect to all other reference systems. Concerning the
Hamiltonian expressions (\ref{3.1}), (\ref{3.5}) and (\ref{3.11}) obtained
above, one observes that all of them depend strongly on the vacuum potential
field function $\bar{W}:\mathbb{M}^{4}\mathbb{\rightarrow R},$ thereby
dissolving the mass problem of the classical energy expression, before
pointed out \cite{Br} by L. Brillouin. It is necessary here to mention that
subject to the canonical Dirac type quantization procedure it can be applied
only to the corresponding dynamical field systems considered with respect to
their proper "rest" reference systems.

\begin{remark}
\textit{{Some comments can be also made concerning the classical relativity
principle. Namely, we have obtained our results completely without using the
Lorentz transformations of reference systems but only the natural notion of
the "rest" reference system and suitable its parametrization with respect to
any other moving reference systems. It looks reasonable since, in reality,
the true state changes of a moving charged particle $q$ are exactly realized
only with respect to its proper "rest" reference system. Thereby, the only
question, still here left open, is that about the physical justification of
the corresponding relationship between time parameters of moving and "rest"
reference systems.} }
\end{remark}

This relationship, being accepted throughout this work, looks as
\begin{equation}
d\tau =dt(1-u^{2})^{1/2},  \label{4.1}
\end{equation}%
where $u:=dr/dt\in \mathbb{R}^{3}$ is the velocity vector with which the
"rest" reference system $\mathcal{K}_{r}$ moves with respect to other
arbitrarily chosen reference system $\mathcal{K}.$ The expression (\ref{4.1}%
) means, in particular, that there holds the equality
\begin{equation}
dt^{2}-dr^{2}=d\tau ^{2},  \label{4.2}
\end{equation}%
which exactly coincides with the classical infinitesimal Lorentz invariant.
Its appearance is, evidently, not casual here, since all our dynamical
vacuum field equations were successively derived \cite{BPT,BPT1} \ from the
governing equations on the vacuum potential field function $W:\mathbb{M}^{4}%
\mathbb{\rightarrow R}$ in the form
\begin{equation}
\partial ^{2}W/\partial t^{2}-\nabla ^{2}W=\rho ,\text{ }\partial W/\partial
t+\nabla (vW)=0,\text{ }\partial \rho /\partial t+\nabla (v\rho )=0,
\label{4.3}
\end{equation}%
being \emph{a priori} Lorentz invariant, where we denoted by $\rho \in $ $%
\mathbb{R} $ the charge density and by $v:=dr/dt$ the suitable local
velocity of the vacuum field potential changes. Thereby, the dynamical
infinitesimal Lorentz invariant (\ref{4.2}) reflects this intrinsic
structure of equations (\ref{4.3}). Being rewritten in the following
nonstandard Euclidean form:%
\begin{equation}
dt^{2}=d\tau ^{2}+dr^{2}  \label{4.4}
\end{equation}%
it gives rise to a completely other time relationship between reference
systems $\mathcal{K}$ and $\mathcal{K}_{r}:$%
\begin{equation}
dt=d\tau (1+\dot{r}^{2})^{1/2},  \label{4.5}
\end{equation}%
where, as earlier, we denoted by $\dot{r}:=dr/d\tau $ the related particle
velocity with respect to the "rest" reference system. Thus, we observe that
all our Lagrangian analysis completed in Section 2 is based on the
corresponding functional expressions written in these "Euclidean" space-time
coordinates and with respect to which the least action principle was
applied. So, we see that there exist two alternatives - the first is to
apply the least action principle to the corresponding Lagrangian functions
expressed in the Minkovski type space-time variables with respect to an
arbitrary chosen reference system $\mathcal{K},$ and the second is to apply
the least action principle to the corresponding Lagrangian functions
expressed in the space-time Euclidean type variables with respect to the
"rest" reference system $\mathcal{K}_{r}.$

As a slightly amusing but exciting inference, following from our analysis in
this work, is the fact that all of classical special relativity results,
related with electrodynamics of charged point particles, can be obtained
one-to-one making use of our new definitions of the dynamical particle mass
and the least action principle with respect to the associated Euclidean type
space-time variables parameterizing the "rest" reference system.

An additional remark is here needed concerning the quantization procedure of
proposed electrodynamics models. If the dynamical vacuum field equations are
expressed in the canonical Hamiltonian form, only technical problems left to
quantize them and obtain the corresponding Schr\"{o}dinger type evolution
equations in suitable Hilbert spaces of quantum states. There exists still
another important inference from the approach devised in this work,
consisting in complete lost of the essence of the well known Einsteinian
equivalence principle \cite{LL,Pa,Fe1,Fe,Kl}, becoming superfluous for our
vacuum field theory of electromagnetism and gravity.

Based on the canonical Hamiltonian formalism devised in this work,
concerning the alternative charged point particle electrodynamics models, we
succeeded in treating their Dirac type quantization. The obtained results
were compared with classical ones, but the physically motivated choice of a
true model is left for the future studies. Another important aspect of the
developed vacuum field theory no-geometry approach to combining the
electrodynamics with the gravity consists in singling out the decisive role
of the related "rest" \ reference system $\mathcal{K}_{r}.$ Namely, with
respect to the "rest" reference \ system evolution parameter $\tau \in
\mathbb{R}$ all of our electrodynamics models allow both the Lagrangian and
Hamiltonian formulations suitable for the canonical quantization. The
physical nature of this fact remains, by now, not enough understood. There is, by
now \cite{Pa,LL,Kl,Lo,Lo1}, no physically reasonable
explanation of this decisive role of the "rest" \ reference system, except of the very
interesting  reasonings by R. Feynman  who argued in \cite{Fe1} that the
relativistic expression  for the classical Lorentz force  (\ref{1.11}) has
physical sense only with respect to \ the "rest" reference system variables $%
(r,\tau )\in \mathbb{E}^{4}.$ In the sequel of our work we plan to analyze the quantization scheme in more detail
and make a step toward the vacuum quantum field theory of infinite many
particle systems.

\section{Acknowledgments}

The Authors are cordially thankful to the Abdus Salam International Centre
for Theoretical Physics in Trieste, Italy, for the hospitality during their
research 2007-2008 scholarships. A.P. is, especially, grateful to P.I. Holod
(Kyiv, UKMA), J.M. Stakhira (Lviv, NUL), U. Taneri (Cyprus, EMU), J. S{\l }awianowski (Warsaw, IPPT),
Z. Peradzy{\'{n}}ski (Warsaw, UW) and M. B{\l }aszak (Pozna\'{n}, UP) for
fruitful discussions, useful comments and remarks. Last but not least thanks
go to academician Prof. A.A. Logunov for his interest to the work, as well
to Mrs. Dilys Grilli (Trieste, Publications office, ICTP) and Natalia K.
Prykarpatska for professional help in preparing the manuscript for
publication.

\end{document}